\documentclass[useAMS,usenatbib,twocolumn]{mn2e}
\usepackage{graphicx}
\usepackage{dcolumn}   
\usepackage{bm}        
\usepackage{amssymb}   
\usepackage{color,soul}

\def\bq{\begin{equation}}
\def\eq{\end{equation}}
\def\bqy{\begin{eqnarray}}
\def\eqy{\end{eqnarray}}

\def\ga{\gamma}

\def\la{\lambda}

\begin{document}

\title[Unified Dynamo mechanism]{Modeling astrophysical outflows via the unified Dynamo-Reverse Dynamo mechanism}
\author[M. Lingam and S.M. Mahajan]{Manasvi Lingam$^{1}$\thanks{E-mail:
manasvi@physics.utexas.edu} and Swadesh M. Mahajan$^{1}$\thanks{E-mail:
mahajan@mail.utexas.edu}\\
$^{1}$Institute for Fusion Studies, The University of Texas, Austin, TX 78712, USA}

\pagerange{\pageref{firstpage}--\pageref{lastpage}} \pubyear{2014}

\maketitle

\label{firstpage}

\date{}

\begin{abstract}
The unified Dynamo-Reverse Dynamo (Dy-RDy) mechanism, capable of simultaneously generating large scale outflows and magnetic fields from an ambient microscopic reservoir, is explored in a broad astrophysical context. The Dy-RDy mechanism is derived via Hall magnetohydrodynamics, which unifies the evolution of magnetic field and fluid vorticity. It also introduces an intrinsic length scale, the ion skin depth, allowing for the proper normalization and categorization of microscopic and macroscopic scales. The large scale Alfv\'en Mach number $\mathcal{M}_{A}$, defining the relative ``abundance'' of the flow field to the magnetic field is shown to be tied to a microscopic scale length that reflects the characteristics of the ambient short scale reservoir. The dynamo (Dy), preferentially producing the large scale magnetic field, is the dominant mode when the ambient turbulence is mostly kinetic, while the outflow producing reverse dynamo (RDy) is the principal manifestation of a magnetically dominated turbulent reservoir. It is conjectured that an efficient RDy may be the source of many  observed astrophysical outflows that have $\mathcal{M}_{A} \gg 1$. 
 \end{abstract}


\section{Introduction}
Astrophysical jets are universal: they appear in myriad forms, exhibiting significant variations in luminosity, collimation and velocity. Relativistic jets have been observed in the context of Active Galactic Nuclei (AGNs), micro-quasars and Gamma Ray Bursts (GRBs). Smaller-scale non-relativistic versions have been observed in a much wider array of systems such as low-mass Young Stellar Objects (YSOs), massive X-ray binary systems and Protoplanetary Nebulae (PPNe). Although the word ``jets'' suggests a highly collimated and narrow outflow, diffuse emanations, often termed ``outflows'', have also been detected in stars in their final stages of evolution. In addition, our Sun exhibits jets and flares, albeit ones that are much smaller than the one mentioned above.  In this Letter, we shall refer to jets and their (relatively) diffuse cousins as ``outflows''. A comprehensive discussion of outflows in astrophysical systems is found in \citet{BP84,L85,MC93,F98,L99,MR99,Ly03,D05,B10}.

Given the diversity of their occurrence and the vastly different parameters that these outflows exhibit, the quest for a broad underlying mechanism that can power these outflows has proven elusive although several seminal works in the 1970s and early 1980s \citep{BR76,L76,BZ77,BK79,BP82} did highlight the importance of magnetic fields in the generation of these outflows. Subsequent works introduced the magneto-centrifugal mechanism \citep{S94,KLB99}, magnetic towers \citep{L03} and other mechanisms \citep{LBC91,OL01,L06,C14}, all  mediated by the magnetic field, in their attempts to explain the origin of these outflows.

Most of the preceding analyses used the ideal magnetohydrodynamic (MHD) model, the simplest and highly explored plasma fluid model. However, in the last decade, the importance of the Hall effect in astrophysical systems has been recognized; the Hall effect has been shown to play a crucial role in several scenarios, such as magnetic braking and star formation \citep{KLS11,LKS11}, protostellar discs \citep{BT01} and magnetic field evolution in neutron star crusts \citep{GR92,CAZ04}. This recognition comes along with the realization that MHD may be an inadequate framework for generating the observed outflows \citep{BL01}. 

In order  to explore the simultaneous dynamics of magnetic fields and outflows, one must resort to models more encompassing than MHD. As an illustration, we will analyze incompressible Hall magnetohydrodynamics (HMHD), the simplest ``beyond MHD'' model that captures one crucial two-fluid effect: the Hall current, expressing differential electron-ion motion, introduces a fundamental scale length, the ion skin depth (ideal MHD is scale-free) that provides a convenient fiduciary length in terms of which  one can define `large scale' and `small scale' quantities. HMHD assigns assigns equal weights to, and thereby unifies, the magnetic fields and flows in a natural manner, as demonstrated in Section \ref{SecII}. Qualitatively, this can be understood as a consequence of one of the HMHD equations involving a simultaneous evolution of the magnetic field ${\bf B}$ and the fluid vorticity $\nabla \times {\bf v}$, resulting in the inexorable linking of these two quantities.

In this Letter, we explore what was termed as the `reverse dynamo' mechanism; derived and developed in \citet{M05}. The mechanism employs HMHD, and exhibits potential universality in generating non-relativistic outflows. The original moniker `reverse dynamo' constitutes an inadequate (and incomplete) description of the more encompassing process. Though dynamo based approaches have been around since the 1970s \citep{L76}, with several important developments in the subsequent decades \citep{BH86,Br95,VC01,BF02,Br05,BS05,EB14}, the Dy-RDy approach is conceptually unique as it incorporates the following features:
\begin{itemize}

\item The HMHD is the governing model from the outset (the Hall term is not appended to an essentially MHD calculation). Consequently the magnetic field and the flows are treated on an entirely equal footing. The ``dynamo'' designates the creation of long-scale magnetic fields from a short scale (turbulent) velocity field, and the ``reverse dynamo'' denotes the complementary/opposite phenomenon of generating long-scale flows from a short scale (turbulent) magnetic field. The HMHD based theory is a unified Dy-RDy theory; both flows and magnetic fields emerge simultaneously from a given source of short scale energy be it kinetic or magnetic. 

\item Unlike MHD, the HMHD has an intrinsic length scale. Thus it becomes possible to characterize long and short scales in a well-defined way. This will prove to be crucial in estimating the length scale relevant to, for example, the generated outflows.

\end{itemize}

\section{Dy-RDy: a synopsis} \label{SecII}

We begin with a  short recapitulation of the formalism derived in \citet{M05}. Assuming a constant density (purely for analytic simplicity) and equal temperatures for the electrons and ions $\left(p = p_i + p_e \approx 2nT\right)$, the HMHD yields the following evolution equations for the flow and the magnetic field:
\begin{eqnarray}\label{basiceq}
\frac{\partial {\bf b}}{\partial t} &=& \nabla \times \left[\left({\bf v} - \alpha_0 \nabla \times {\bf b}\right) \times {\bf b}\right], \nonumber \\
\frac{\partial {\bf v}}{\partial t} &=& {\bf v} \times \left(\nabla \times {\bf v}\right) + \left(\nabla \times {\bf b}\right)\times {\bf b} - \nabla \left(p+\frac{v^2}{2}\right),
\end{eqnarray}
wherein $\alpha_0 = \lambda_{i0}/R_0$, with $\lambda_{i0}$ representing the ion skin depth, which serves as the intrinsic length scale characteristic of HMHD. $R_0$ represents a characteristic scale length to be specified shortly hereafter. In (\ref{basiceq}), the magnetic field is normalized to an arbitrary representative magnetic field $B_0$,  the velocity to the corresponding  Alfven velocity $V_{A0}$, the length scale to $R_0$ and the timescale  to $R_0/V_{A}$.  Subsequently, we choose $R_0$ to be the ion skin depth, as we are interested in the underlying microscopic physics. Thus, we observe that $\lambda_{i0}$ \emph{will no longer appear explicitly}, i.e, $\alpha_0=1$. With such a choice of units, one may re-express (\ref{basiceq}) as a pair of vorticity-like equations \citep{MY98}:
\begin{eqnarray} \label{VortEvol}
&& \frac{\partial {\boldsymbol\Omega}_j}{\partial t} - \nabla \times \left({\bf v}_j \times {\boldsymbol\Omega}_j\right) = 0, \nonumber 
\end{eqnarray}
where the vorticities ${\boldsymbol\Omega}_j$ and the associated velocities ${\bf v}_j$ are
\begin{eqnarray} \label{VortMomDef}
&& {\boldsymbol\Omega}_1 = {\bf B}, \quad {\bf v}_1 = {\bf v} - \nabla \times {\bf B}, \nonumber \\
&& {\boldsymbol\Omega}_2 = {\bf B} + \nabla \times {\bf v}, \quad {\bf V}_2 = {\bf v}. \nonumber
\end{eqnarray}
The structure of the above equations is radically different from that of ideal MHD, and is also indicative why ${\bf B}$ and ${\bf v}$ $\left(\nabla \times {\bf v}\,\mathrm{to\,be\,precise}\right)$ are generated in tandem. 

Notice that the governing equation for the canonical vorticity ${\boldsymbol\Omega}_2$ treats the magnetic field and the vorticity on par. There are two conserved helicities in the system: the magnetic helicity $\int_D d^3x\,{\bf A}\cdot{\bf B}$ and the canonical helicity $\int_D d^3x\,\left({\bf A}+{\bf v}\right)\cdot\left(\nabla \times {\bf v} + {\bf B}\right)$; the latter is essentially the ion helicity reflecting the (finite) ion inertia. Unlike in ideal MHD, the cross helicity, $\int_D d^3x\,{\bf v}\cdot{\bf B}$, is no longer conserved.

The velocity and magnetic fields are decomposed into the equilibrium seed fields (${\bf v_0}$ and ${\bf b_0}$) and the fluctuations; the latter, in turn, comprising of the macroscopic (${\bf U}$ and ${\bf H}$) and microscopic (${\bf \tilde{v}}$ and ${\bf \tilde{b}}$) components: 
\begin{eqnarray}
{\bf b} &=& {\bf H + \tilde{b} + b_0}, \nonumber \\
{\bf v} &=& {\bf U + \tilde{v} + v_0}.
\end{eqnarray}
It is important to realize that the fluctuations are comprised of both macroscopic and microscopic components. 
The former yield non-zero expressions upon carrying out a suitable ensemble or spatial averaging, whilst linear combinations of the latter yield no contributions. Over a microscopic scale, we also assume that the large scale fluctuations do not vary significantly, i.e. that their derivatives vanish. We refer the reader to \citet{MGM02,MGM03a,MGM03b,M05} for a detailed discussion of the closure scheme, and its accompanying assumptions.

The equilibrium fields are the energy reservoir for driving the fluctuations. A comment regarding the length scales is in order.  The only intrinsic scale is defined by the normalizing length, the ion skin depth. Any macroscopic length scale of interest is way bigger than $\lambda_{i0}$, but there is some latitude in what could be called the microscopic length. The latter (microscopic length) is taken to be any length that is on the order of, or smaller than, the ion skin depth $\lambda_{i0}$ (or of order unity and less in normalized units). The equilibrium fields are small scale (microscopic), have been produced by some microscopic process, undergone saturation, and have built up an energy reservoir that  will drive the large scale (macroscopic and observable) and small scale (microscopic) fluctuations. 

Given that one has little information about the \emph{equilibrium} fields, one makes the most `natural' choice in the context of Hall MHD; the double Beltrami equilibria, described in \citet{MY98}. The double Beltrami equilibria are the solutions of
\begin{equation}
\frac{{\bf b_0}}{a} + \nabla \times {\bf b_0} = {\bf v_0},\quad {\bf b_0} + \nabla \times {\bf v_0} = d {\bf v_0},
\end{equation}
which, along with the Bernoulli condition \\ $\nabla \left(p+\frac{v^2}{2}\right)=0$ constitute the stationary solutions of (\ref{basiceq}). The constants $a$ and $d$, set by the helicities of the system \citep{M05}, define two (dimensionless) inverse scale lengths $\lambda_{\pm} = \frac{1}{2} \left[\left(d-a^{-1}\right) \pm \sqrt{\left(d+a^{-1}\right)^2-4} \right]$. Appropriate tuning of $a$ and $d$ can lead to vastly separated roots, $\lambda_{+} \equiv \Lambda$ and $\lambda_{-} \equiv \zeta$; the former (latter) represents the microscopic (macroscopic) \emph{inverse} scale length. Such a choice is physically meaningful - the astrophysical systems are macroscopic, but their underlying physics  may have a substantial microscopic component. In fact, we could assume that the ambient fields are entirely microscopic; this is again a rational assumption since the underlying reservoir that generates observable (macroscopic) phenomena is, indeed, presumed to be  microscopic.

With these assumptions and a great deal of algebra \citep{M05}, one derives the final equations for the macroscopic fluctuations:
\begin{equation} \label{HUGov}
\ddot{\bf H} = -r \left(\nabla \times {\bf H}\right), \quad
\ddot{\bf U} = \nabla \times \left(s{\bf U} - q {\bf H}\right)
\end{equation}
where $r$, $s$ and $q$, given by
\begin{eqnarray} \label{qrs}
q &=& \Lambda^2 \frac{b_0^2}{6}, \quad r = \Lambda \frac{b_0^2}{3} \left(1-\Lambda a^{-1} - a^{-2}\right), \nonumber \\
s &=& \Lambda \frac{b_0^2}{6} \left[\left(\Lambda + a^{-1}\right)^2 - 1\right],
\end{eqnarray}
are determined by $b^2_0$, representing the normalized ambient magnetic energy, and by the scale lengths set by the microscopic helicities; the entire macroscopic dynamics is controlled by the ambient microscopic dynamics. By manipulating the equations in (\ref{HUGov}), we find that
\begin{equation} \label{UHrel}
{\bf U} = \frac{q}{s+r} {\bf H}.
\end{equation}
Thus, by solving the first equation in (\ref{HUGov}), we can fully determine $\bf H$ and  $\bf U$. The details of the solution can be found in \citet{M05}.

\section{Dy-RDy in astrophysical systems} \label{SecIII}

\begin{table*}
\begin{minipage}{126mm}
\caption{The Alfv\'en Mach number for astrophysical outflows}
\label{Tab1}
\begin{tabular}{|c|c|c|c|c|c|}
\hline 
System & $\rho$$\left(g/cm^{3}\right)$ & $H$$\left(G\right)$ & $U$$\left(cm/s\right)$ & $\mathcal{M}_A$ & Reference\tabularnewline
\hline 
\hline 
GRBs & $10^{16}$ & $10^{15}$ & $10^{10}$ & $10^{3}$ & \citet{B10}\tabularnewline
\hline 
Microquasars & $10^{16}$ & $10^{10}$ & $10^{10}$ & $10^{8}$ & \citet{B10}\tabularnewline
\hline 
Radio pulsars & $10^{15}$ & $10^{12}$ & $10^{10}$ & $10^{6}$ & \citet{B10}\tabularnewline
\hline 
YSOs & $10$ & $10^{3}$ & $10^{7}$ & $10^{5}$ & \citet{B10}\tabularnewline
\hline 
PPNe & $1$ & $10^{2}$ & $10^{7}$ & $10^{5}$ & \citet{H07,AR14}\tabularnewline
\hline 
\end{tabular}
\medskip

The table illustrates the values of the Alfv\'en Mach number $\mathcal{M}_A$ for astrophysical systems where jets are observed. In the case of Protoplanetary Nebulae (PPNe), the outflows are modeled as originating from the central star, which possesses specifications akin to the Sun. An upper bound on the large scale magnetic field in PPNe was obtained in \citet{AR14}, and the typical jet velocities in PPNe were taken from \citet{H07}. The rest of the parameters are presented in Table 1 of \citet{B10}.
\end{minipage}
\end{table*}

Equation \ref{UHrel} is the most potent and revealing statement of the Unified dynamo - the simultaneous generation of large scale magnetic fields and flows out of a short-scale reservoir of kinetic and magnetic energy. The dominant dynamo $|\bf{H}|\gg|\bf{U}|$ or the dominant reverse dynamo $|\bf{U}|\gg|\bf{H}|$ state are simply the two limiting cases of the same mechanism. 

What mix of  $|\bf{H}|$ and $|\bf{U}|$ will finally emerge depends upon the coefficients $q$, $s$,and $r$, which are in turn determined by $a$ and $d$. The latter duo are also manifest in the scale lengths, namely $\Lambda^{-1}$ and $\zeta^{-1}$. The helicity measures, in turn, reflect the initial state of the system, in particular, the ratio of the ambient kinetic energy to the ambient magnetic energy. No matter what their ratio, the flows and the magnetic fields are generated together; the HMHD is a theory of the unified Dynamo-Reverse Dynamo (Dy-RDy) process.

Simple algebra reveals that for $a \sim d \ll 1$, the short scale reservoir is mostly magnetic, ${\bf v_0} \sim a {\bf b_0} \ll {\bf b_0}$ leading to strong (compared to magnetic fields) macroscopic flows ${\bf U} \sim \Lambda {\bf H} \gg {\bf H}$. This limit pertains only to those systems where the observed flows are super-Alfv\'enic. The standard dynamo scenario pertains to the opposite case: $a \sim d \gg 1$, implying the microscopic relation ${\bf v_0} \sim a {\bf b_0} \gg {\bf b_0}$, and the macroscopic relation ${\bf H}\gg {\bf U}$. 

This deterministic relationship between the energy mix in the short scale reservoir, and the consequent long scale observables is one of the major results of the paper.  The theory has provided a probe into the earlier era microphysics; measuring the long range fields and flows yields information about the source that caused them.

Let us explore a little further the regime where  RDy part of Dy-RDy will be dominant. We will examine the relevant relation ${\bf U} \sim \Lambda {\bf H}$ in physical units. Since  $\bf{H}$ and $\bf{U}$ are normalized by $B_0$ and $V_{A0} = B_0/\sqrt{4\pi m_i n}$ (with constant $n$), we find that
\begin{equation} \label{MachL}
\Lambda \sim \frac{\bar{U}}{\bar{H}/\sqrt{4\pi m_i n}} \sim \frac{\bar{U}}{\bar{\mathcal{V}}_{A}} \sim \mathcal{M}_{A}.
\end{equation}
Hence, the inverse (dimensionless) microscopic scale length $\Lambda$ turns out to be approximately equal to the large scale Alfv\'en Mach Number $\mathcal{M}_{A}$, where $\bar{U}$ and $\bar{\mathcal{V}}_{A}$  are, respectively, the (dimensional) large scale flow and Alfv\'en velocities. Equation (\ref{MachL}) captures the very essence of this analysis, as it directly ties together the microscopic and macroscopic physics. This is also seen in (\ref{UHrel}), as it relates $q/(s+r)$ (microscopic) with $U/H$ (macroscopic). In our subsequent discussion, we drop the overbars and note that all quantities on the RHS of (\ref{MachL}) are large scale and dimensional. 

The reverse dynamo mechanism, favored by the ordering $a \sim d \ll 1$ ($\Lambda \gg 1$), will operate exactly in those regimes where the observed $\mathcal{M}_A$ is much greater than unity. Equivalently, if the observed large scale flows are highly super-Alfv\'enic, then it is quite likely that they are generated via the reverse dynamo mechanism. The characteristic Mach number $\mathcal{M}_A$ of a given outflow is an index of the relative efficiency of the RDy and Dy mechanisms, which in turn, is a mirror of the constitution of the ambient state - what mixture of magnetic to kinetic turbulence it is endowed with.

\begin{figure}
\quad\quad\quad \includegraphics[width=5.68cm]{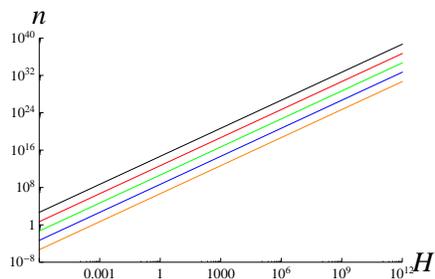} \\
\caption{The figure depicts the permitted values of $n$ and $H$ for the reverse dynamo mechanism. For all the curves, $\Lambda=10$ has been chosen. The black, red, green, blue and orange curves correspond to the cases with $U=1\,,10\,,10^2\,,10^3\,,10^4$ km/s respectively. Note that this is a log-log plot.}
\label{fig1}
\end{figure}

Let us now use (\ref{MachL}) to probe the range of densities and magnetic fields for which the RDy is likely to be important. Choosing $\Lambda \gtrsim 10$ (which loosely satisfies $\Lambda \gg 1$) and different values of $U$, we can calculate $\mathcal{V}_A$, and the corresponding values of $H$ and $n$. The results are presented in Fig. \ref{fig1}. For the current non-relativistic treatment, we keep $U$ bounded below $10^4$ km/s. Each of the curves in Fig. \ref{fig1} corresponds to a fixed value of $U$, and allows us to probe the ranges of $n$ and $B$ for which the reverse dynamo may be operational. In the regions lying above and on the curve, RDy will dominate and generate large scale outflows, while in the regions below the curves, the RDy will be sub-dominant.  

In Table \ref{Tab1}, the computed Alfv\'en Mach numbers are displayed for a variety of astrophysical systems. The computations should be accepted with the following caveats:
\begin{itemize}
\item The magnetic fields $\left(H\right)$ in the large scale outflows cannot be easily measured, and hence we have chosen to use the magnetic fields present in the `engine' (driving the outflows) as a substitute. However, since most of the values of $\mathcal{M}_A$ are very high, it is reasonable to suppose that reducing the magnetic field by several orders of magnitude may still plant the system firmly in the regime where the reverse dynamo mechanism determines the outcome.

\item The magnetic fields in both the `engine' and the outflows are both very complex, and the Alfv\'en Mach number serves only as a simplified criterion for evaluating the validity of the reverse dynamo mechanism. In addition, it is known that the winds in GRBs, microquasars and pulsars are relativistic, and a fully consistent treatment must take such effects into account. We plan to present a relativistic version in a forthcoming paper. 
\end{itemize}

The Alfv\'en Mach numbers, displayed in Table \ref{Tab1},  are all very high - much greater than unity. The simplest criterion that such outflows may originate in a RDy mechanism is amply satisfied. It must be borne in mind that this does not necessarily imply the operation of the RDy mechanism. Nevertheless, this simple criterion can help us probe the nature of the underlying source that drives large scale outflows and magnetic fields. An example in contrast is the solar wind, a highly sub-Alfv\'enic outflow; the RDy is not likely to be the primary driver. However, we note that there are regions in the chromosphere where the RDy has been shown to play a key role \citep{M02}.

Lastly, we note that the HMHD is a theory of unified Dy-RDy mechanism - the general theory is just as valid in scenarios where $\mathcal{M}_A \lesssim 1$. Even in this sub-Alfv\'enic regime, the flows, along with the dominant magnetic fields, will continue to amplify as long as there is turbulent energy to drive them. Both fields grow at the rate determined by the dispersion relation, $\omega^2 = - \left|r\right|k$ \citep{M05} with $r$ given by (\ref{qrs}). One expects the outflow level to be boosted over time, albeit in a sub-Alfv\'enic setting, until it is compatible with observations. 

\section{Discussion and conclusion} \label{SecIV}
In this Letter, we have explored the unified Dynamo-Reverse Dynamo mechanism within an HMHD model. The RDy (Dy), operates preferentially when the magnetic component (the kinetic component) accounts for the bulk of the short scale energy reservoir. The end product of dominant RDy (Dy) is a super Alfv\'enic (sub Alfv\'enic) outflow, where the Alfv\'en Mach number obeys $\mathcal{M}_A \gg 1$ $\left(\mathcal{M}_A \ll 1\right)$. Consequently, if the observed large scale outflows were to satisfy $\mathcal{M}_A \gg 1$ $\left(\mathcal{M}_A \ll 1\right)$, they may indicate the operation of a reverse dynamo (dynamo) mechanism. A knowledge of $\mathcal{M}_A$ as per this model can serve as a probe of the ``content'' of the underlying ambient reservoir. 

The unified Dy-RDy mechanism is characterized by a striking reciprocity between the micro and the macro scales: the ambient (microscopic) Alfv\'en Mach number, denoted by $\cal{M}_{\mu A}$,  precisely equals $\left({\cal{M}_{A}}\right)^{-1}$ - the \emph{inverse} Alfv\'en Mach number of the large scale (macroscopic) outflow. 

We have also found that the reverse dynamo mechanism may operate in a wide variety of astrophysical objects whose ambient density $n$ and the large scale magnetic fields $H$ lie in the ranges delineated in Fig. \ref{fig1} and Table \ref{Tab1}. We do not claim that the RDy mechanism is the sole source of super-Alfv\'enic outflows. Instead, we emphasize the importance of the large scale Alfv\'en Mach number $\mathcal{M}_A$ as a means of gauging the composition of the ambient magnetic and kinetic energies of the reservoir. However, our analysis is not exhaustive;  some of the inherent limitations were outlined earlier.

We must stress that the Hall term, which introduces an intrinsic micro-scale, and thereby opens up the possibility of physics at disparate macros and micro scales, is an essential enabler of the unified Dy-RDy theory. This step, however, represents only the tip of the iceberg as Hall MHD, itself, is the simplest of the extended MHD models. A natural extension of the formalism presented in this paper involves the construction of an extended MHD and/or two-fluid unified Dy-RDy that can capture the non-ideal MHD effects in a more direct manner. To understand relativistic flows, we must resort to relativistic MHD \citep{L67} or the coupled relativistic magnetofluid model \citep{M03}. For incompressible Hall MHD, we hypothesize that the results are likely to remain similar under ${\bf v} \rightarrow  \gamma {\bf v}$ in the relativistic regime. Lastly, we note that the assumption of homogeneity can also be relaxed without difficulty, and the ensuing model is still easily solvable via computational means. 

The unified Dy-RDy mechanism presented herein (and/or one of its several suggested extensions) constitutes a very viable candidate, as well as a strong indicator, for generating large scale outflows in a wide range of astrophysical systems. A more detailed picture of the model's strengths and limitations are likely to emerge only via further numerical simulations.

\section*{Acknowledgments}
SMM's research was supported by the U.S. Dept. of Energy Grant DE-FG02-04ER-54742. ML's research was supported by the U.S. Dept. of Energy Contract DE-FG05-80ET-53088.

\label{lastpage}


\begin{thebibliography}{}

\bibitem[Asensio Ramos et al.(2014)]{AR14} Asensio Ramos, A., Mart\'inez Gonz\'alez, M. J., Manso Sainz, R., Corradi R. L. M., 
    \& Leone, F.  2014, ApJ, 787, 111

\bibitem[Balbus \& Terquem(2001)]{BT01} Balbus, S. A.,
    \& Terquem, C.  2001, ApJ, 552, 235
    
\bibitem[Beskin(2010)]{B10}
Beskin, V. S. 2010, PhyU, 53, 1199

\bibitem[Bhattacharjee \& Hameiri(1986)]{BH86} Bhattacharjee, A.
    \& Hameiri, E. 1986, Phys. Rev. Lett., 56, 206

\bibitem[Bisnovatyi-Kogan \& Ruzmaikin(1976)]{BR76} Bisnovatyi-Kogan, G. S.,
    \& Ruzmaikin, A. A.  1976, Ap\&SS, 42, 401
    
\bibitem[Bisnovatyi-Kogan \& Lovelace(2001)]{BL01} Bisnovatyi-Kogan, G. S.,
    \& Lovelace, R. V. E.  2001, NewA Rev., 45, 663
    
\bibitem[Blackman \& Field(2002)]{BF02} Blackman, E. G.,
    \& Field, G. B. 2002, Phys. Rev. Lett., 89, 265007
    
\bibitem[Blandford \& Znajek(1977)]{BZ77} Blandford, R. D.,
    \& Znajek, R. L.  1977, MNRAS, 179, 433
    
\bibitem[Blandford \& K{\"o}nigl(1979)]{BK79} Blandford, R. D.,
    \& K{\"o}nigl, A.  1979, ApJ, 232, 34
    
\bibitem[Blandford \& Payne(1982)]{BP82} Blandford, R. D.,
    \& Payne, D. G.  1982, MNRAS, 199, 883
    
\bibitem[Brandenburg (2005)]{Br05} Brandenburg, A.  2005, ApJ, 625, 539
    
\bibitem[Brandenburg et al.(1995)]{Br95} Brandenburg, A., Nordlund, A., Stein, R. F., 
    \& Torkelsson, U.  2005, ApJ, 446, 741
    
\bibitem[Brandenburg \& Subramanian(2005)]{BS05} Brandenburg, A.,
    \& Subramanian, K.  2005, PhR, 417, 1

    
\bibitem[Bridle \& Perley(1984)]{BP84} Bridle, A. H., 
    \& Perley, R. A. 1984, ARA\&A, 22, 319
    
\bibitem[Colgate et al.(2014)]{C14} Colgate, S. A., Fowler, K. T., Li, H., \& Pino, J. 2014, ApJ, 789, 144

\bibitem[Cumming, Arras \& Zweibel(2004)]{CAZ04} Cumming, A., Arras, P., \& Zweibel, E. 2004, ApJ, 609, 999

\bibitem[de Gouveia Dal Pino(2005)]{D05}
de Gouveia Dal Pino, E. M.  2005, Adv. Space Res., 35, 908

\bibitem[Ebrahimi \& Bhattacharjee(2014)]{EB14} Ebrahimi, F.,
    \& Bhattacharjee, A. 2014, Phys. Rev. Lett., 112, 125003

\bibitem[Ferrari(1998)]{F98}
Ferrari, A. 1998, ARA\&A, 36, 539

\bibitem[Goldreich \& Reisenegger(1992)]{GR92} Goldreich, P., \& Reisenegger, A. 1992, ApJ, 395, 250

\bibitem[Huggins(2007)]{H07}
Huggins, P. J. 2007, ApJ, 663, 342

\bibitem[Krasnopolsky, Li \& Blandford(1999)]{KLB99} Krasnopolsky, R., Li, Z.-Y.,
    \& Blandford, R.  1994, ApJ, 526, 631
    
\bibitem[Krasnopolsky, Li \& Shang(2011)]{KLS11} Krasnopolsky, R., Li, Z.-Y.,
    \& Shang, H.  2011, ApJ, 733, 54
    
\bibitem[Lada(1985)]{L85}
Lada, C. J.  1985, ARA\&A, 23, 267
    
\bibitem[Li et al.(2006)]{L06} Li, H., Lapenta, G., Finn, J. M., Li, S., \& Colgate, S. A. 2006, ApJ, 643, 92

\bibitem[Li, Krasnopolsky \& Shang(2011)]{LKS11} Li, Z.-Y., Krasnopolsky, R., \& Shang, H.  2011, ApJ, 738, 180

\bibitem[Lichnerowicz(1967)]{L67}
Lichnerowicz, A. 1967, Relativistic Hydrodynamics and Magnetohydrodynamics (New York: Benjamin)

\bibitem[Livio(1999)]{L99}
Livio, M.  1999, PhR, 311, 225
    
\bibitem[Lovelace(1976)]{L76}
Lovelace, R. V. E.  1976, Nature, 262, 649
    
 \bibitem[Lovelace, Berk \& Contopoulos(1991)]{LBC91}
Lovelace, R. V. E., Berk, H. L., 
    \& Contopoulos, J.  1991, ApJ, 379, 696
    
\bibitem[Lynden-Bell(2003)]{L03}
Lynden-Bell, D.  2003, MNRAS, 341, 1360

\bibitem[Lyutikov \& Blandford(2003)]{Ly03}
Lyutikov, M., \& Blandford, R.  2003, arXiv:astro-ph/0312347

\bibitem[Mahajan(2003)]{M03} Mahajan, S. M. 2003, Phys. Rev. Lett., 90, 035001

\bibitem[Mahajan \& Yoshida(1998)]{MY98} Mahajan, S. M., \& Yoshida, Z. 1998, Phys. Rev. Lett., 81, 4863

\bibitem[Mahajan et al.(2002)]{M02} Mahajan, S. M., Nikol'skaya, K. I., Shatashvili, N. L., \& Yoshida, Z. 2002, ApJL, 576, L161

 \bibitem[Mininni, G\'omez \& Mahajan(2002)]{MGM02}
Mininni, P. D., G\'omez, D. O., 
    \& Mahajan, S. M.  2002, ApJL, 567, L81

 \bibitem[Mininni, G\'omez \& Mahajan(2003a)]{MGM03a}
Mininni, P. D., G\'omez, D. O., 
    \& Mahajan, S. M.  2003a, ApJ, 584, 1120
    
    \bibitem[Mininni, G\'omez \& Mahajan(2003b)]{MGM03b}
Mininni, P. D., G\'omez, D. O., 
    \& Mahajan, S. M.  2003b, ApJ, 587, 472

\bibitem[Mahajan et al.(2005)]{M05} Mahajan, S. M., Shatashvili, N. L., Mikeladze, S. V., \& Sigua, K. I. 2005, ApJ, 634, 419

\bibitem[Masson \& Chernin(1993)]{MC93}
Masson, C. R.,
    \& Chernin, L. M.  1993, ApJ, 414, 230
    
\bibitem[Mirabel \& Rodr{\'i}guez(1999)]{MR99}
Mirabel, I. F.,
    \& Rodr{\'i}guez, L. F.  1999, ARA\&A, 37, 409

\bibitem[Ogilvie \& Livio(2001)]{OL01} Ogilvie, G. I.,
    \& Livio, M.  2001, ApJ, 553, 158
    
\bibitem[Shu et al.(1999)]{S94} Shu, F., Najita, J., Ostriker, E., Wilkin, F., Ruden, S. \& Lizano, S. 1994, ApJ, 429, 781

\bibitem[Vishniac \& Cho(2001)]{VC01} Vishniac, T. E.,
    \& Cho, J.  2001, ApJ, 550, 752


\end{thebibliography}
\end{document}